\documentclass[conference]{IEEEtran}

\usepackage{graphicx,epsfig}
\usepackage[noadjust]{cite}
\usepackage{mcite}
\usepackage{amsfonts,helvet}
\usepackage{fancyhdr}
\usepackage{threeparttable}
\usepackage{epsf,epsfig}
\usepackage{amsthm}
\usepackage{amsmath}
\usepackage{siunitx}
\usepackage{amssymb}
\usepackage{booktabs}
\usepackage{dsfont}
\usepackage{subfigure}
\usepackage{color}
\usepackage[linesnumbered,ruled,noend]{algorithm2e}
\usepackage{algpseudocode}
\usepackage{algcompatible}
\usepackage{enumerate}
\usepackage{gensymb}
\usepackage{cancel}
\usepackage{bbm}
\usepackage{graphicx,subfigure}
\usepackage{graphicx}
\usepackage{array}
\usepackage{mathtools}
\usepackage{textcomp}
\usepackage{siunitx}
\newcolumntype{P}[1]{>{\centering\arraybackslash}p{#1}}

\usepackage{eucal}

\begin{document}
\bstctlcite{IEEEexample:BSTcontrol}

\title{
    Cooperative Dynamic Spectrum Access for Cell-Free MISO Networks Using Spectrum Consumption Models
}

\author{
    Jiwon Sung$^\dagger$,
    Seungmin Choi$^\dagger$,
    Abhiram R. Gorle$^\dagger$,
    Atul A. Salvekar$^\dagger$,
    Igor Kadota$^\ddagger$,
    John M. Cioffi$^\dagger$
    \\
    {\normalsize
        $^\dagger$Department of Electrical Engineering, Stanford University, Stanford, CA, USA
    }
    \\
    {\normalsize
        $^\ddagger$Department of Electrical and Computer Engineering, Northwestern University, Evanston, USA
    }
    \\
    {\normalsize E-mail: 
    \{jwsung, smchoi, abhiramg, salvekar, cioffi\}@stanford.edu,
    \ kadota@northwestern.edu
    }
}

\maketitle \setcounter{page}{1}

\begin{abstract}
    This paper incorporates uniform planar arrays into a cooperative dynamic spectrum access (DSA) mechanism based on the IEEE 1900.5.2 spectrum consumption model (SCM).
    Unlike prior work on SCM-based DSA that assume a fixed radiation pattern, the proposed DSA mechanism treats the radiation pattern as a design variable and jointly designs the precoder and spectrum allocation.
    The system model is a cell-free downlink, multiuser, multiple-input single-output network, an architecture that suits the dense indoor deployments of large-scale networks such as online retailers' fulfillment centers.
    The main objective is to reduce spectrum use while meeting a target signal-to-interference-plus-noise ratio at every receiver.
    To this end, this paper proposes a leakage-aware iterative water-filling precoder using uplink-downlink reciprocity, a power-control method that further encourages spectrum sharing, and a sequential admission algorithm that allocates spectrum to base stations based on the SCM compatibility test.
    Simulations in an indoor warehouse setting show that the proposed DSA mechanism better utilizes the spatial dimension for spectrum sharing.
\end{abstract}

\section{Introduction}

Spectrum sharing has become critical for meeting the explosive demand for spectrum resources with the ever-increasing amount of Internet of Things (IoT) devices that are being deployed each year~\cite{zhang2018spectrum}.
Spectrum sharing is especially important for time-sensitive collaborative multi-agent systems, such as online retailers' vast fulfillment warehouse centers in the US~\cite{tripathi2023wiswarm}.
In such large-scale scenarios, dynamic spectrum access (DSA)~\cite{netalkar2025scalable, bhardwaj2016enhanced, yu2022spectrum, tamim2025cooperative} can efficiently use the limited available spectrum resources and enable coexistence between radio frequency (RF) devices.
Although most prior work on DSA employs omnidirectional antennas~\cite{agiwal2016next}, there is some work that uses directional antennas~\cite{bhardwaj2016enhanced, yu2022spectrum, tamim2025cooperative}.
Recent work, in particular~\cite{tamim2025cooperative}, uses a measured radiation pattern of a 28 GHz horn antenna as well as the radiation-pattern envelopes that the European Telecommunications Standards Institute (ETSI) specifies for fixed radio systems in the $24$ to $30$ GHz range.
Utilizing the spatial dimension enables the DSA mechanism to reduce further the interference inflicted on other receivers, resulting in better spectrum sharing.
In addition, \cite{tamim2025cooperative} also considers cooperative DSA, where RF devices exchange information to select a spectrum configuration (e.g., carrier frequency and power) that avoids harmful interference.
It is based on the IEEE 1900.5.2 standard~\cite{scm}, a recent IEEE standardization effort on DSA, commonly referred to as spectrum consumption models (SCMs).

Another recent work on SCM-based DSA identifies phased array antennas as a future research direction~\cite{netalkar2025scalable}.
Phased arrays can steer the beams electronically by adjusting their elements' complex weights, unlike the directional antennas in~\cite{tamim2025cooperative} that have a fixed radiation pattern that cannot be adaptively adjusted to changing environments.
Digital beamforming goes one step further and can form multiple beams at once, which enables a transmitter to serve a group of users over the same time and frequency resources by separating them in the spatial domain, instead of the single transmitter-receiver paired links considered in~\cite{netalkar2025scalable, tamim2025cooperative}.
This paper's proposed DSA mechanism extends the SCM standardization efforts to multiuser systems by incorporating uniform planar arrays (UPAs) into the spectrum-sharing algorithm.

This paper studies cooperative DSA for a cell-free downlink, multiuser, multiple-input single-output (MISO) system, an architecture that suits the dense indoor deployments of large-scale networks such as online retailers' fulfillment centers.
Unlike prior work on SCM-based DSA that assumes a fixed radiation pattern, the proposed DSA mechanism treats the radiation pattern as a design variable.
Each base station (BS) declares its spectrum use through an SCM and obtains its carrier frequency based on the IEEE standard's \emph{compatibility test (CT)}.
By jointly designing the precoding weights and spectrum allocation, the proposed DSA mechanism better utilizes the spatial dimension.
This paper's main contributions are
($i$) a system model where each BS's radiation pattern depends on precoding rather than a fixed measured antenna pattern,
($ii$) a leakage-aware iterative water-filling (IWF) precoder derived using uplink-downlink reciprocity,
($iii$) a power-control method that is designed to meet a predefined target signal-to-interference-plus-noise ratio (SINR),
and ($iv$) a sequential admission algorithm that evaluates the CT using IWF and power control, and allocates spectrum to BSs that pass the CT.
Simulations are set in an indoor warehouse setting and the results show the effectiveness of the proposed DSA mechanism in terms of spectrum sharing and spectral efficiency performance.

\subsection{Background on Spectrum Consumption Models}
\label{subsec:scm_primer}

\begin{figure}[!t]
    \centering
    \subfigure[Tx spectrum mask]{%
          \includegraphics[width=0.49\columnwidth]{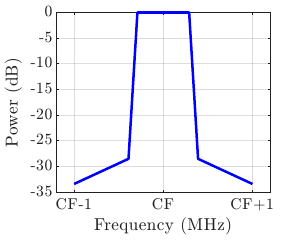}%
          \label{fig:mask_tx}}
    \hfil
    \subfigure[Rx underlay mask]{%
          \includegraphics[width=0.49\columnwidth]{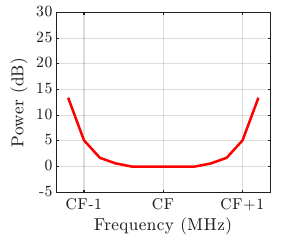}%
          \label{fig:mask_rx}}
    \caption{
        Illustration of the (a) spectrum mask and (b) underlay mask SCM constructs used in the compatibility test, derived from measurements of BPSK modulated signals with a $1$~MHz bandwidth in \protect\cite{stojadinovic2022spectrum}.
    }
    \label{fig:masks}
\end{figure}

An SCM models RF devices' spectral, spatial, and temporal characteristics.
To describe the system, SCMs use a set of up to $11$ \emph{constructs} defined in the IEEE 1900.5.2 standard~\cite{scm}.
These constructs classify SCMs:
($i$) \emph{transmitter models}, which convey the extent and strength of RF emissions from a transmitter,
($ii$) \emph{receiver models}, which convey when an RF receiver is experiencing too much aggregate interference,
and ($iii$) \emph{system and set models}, which are aggregates of transmitter and receiver models.
The SCM used in this paper uses the following four constructs:
\begin{itemize}
    \item \emph{Reference power:} Value that provides a reference power level for transmitter emission or for the allowed interference into a receiver. 
    It serves as the reference power value for the spectrum mask and the underlay mask constructs.
    
    \item \emph{Spectrum mask:} Defines the emissions' relative spectral power density. 
    Fig.~\ref{fig:mask_tx} shows an example.
    This construct is mandatory for transmitter models only.
    
    \item \emph{Underlay mask:} Defines the relative spectral power density of allowed interference. 
    Fig.~\ref{fig:mask_rx} shows an example.
    This construct is mandatory for receiver models only.
    
    
    
    \item \emph{Location:} Specifies an RF device's location.
\end{itemize}
The 1900.5.2 standard requires their use in both transmitter and receiver models unless stated otherwise.

Using the information provided in an SCM, the procedures defined in the 1900.5.2 standard can be used to evaluate whether the spectrum use of multiple transmitters and receivers is compatible.
In particular, the compatibility test indicates whether the spectrum can be shared by two or more RF devices without severe performance degradations caused by interference.

\section{System Model}
\label{sec:sys_model}

The model is a cell-free downlink, multiuser, MISO system where $B$ BSs cooperatively serve $K$ single-antenna users over a shared spectral band.
The BSs are ceiling-mounted at height $h_b$ over a square coverage area $[0,L] \times [0,L]$ and are arranged on a $\sqrt{B}\times\sqrt{B}$ square grid, each placed at its sub-cell's center.
The $K$ users are located at independent, uniformly random positions in the coverage area at a fixed height $h_{\sf u} \ll h_b$.
Each BS has an $N$-element UPA mounted horizontally facing the users underneath.

\subsection{UPA Array Response Vector}

Each BS's UPA has $N=N_x N_y$ elements with half-wavelength spacing $\lambda / 2$.
The vector $(r_{b,x}, r_{b,y}, h_b)$ denotes the location of BS $b$ and $(r_{k,x}, r_{k,y}, h_{\sf u})$ denotes the location of user $k$.
Then, the distance between BS $b$ and user $k$ is
\begin{align}
    d_{b,k} &= \sqrt{(r_{k,x}-r_{b,x})^2+(r_{k,y}-r_{b,y})^2+(h_{\sf u} - h_b)^2},
\end{align}
and the corresponding $x$- and $y$-direction cosines are
\begin{align}
    \label{eq:dir_cos}
    u_{b,k} &= \frac{r_{k,x}-r_{b,x}}{d_{b,k}}, \qquad
    v_{b,k} = \frac{r_{k,y}-r_{b,y}}{d_{b,k}}.
\end{align}
Using \eqref{eq:dir_cos}, the UPA array response \cite{van2002optimum} toward a user is
\begin{align}
    \label{eq:steering}
    \big[\mathbf{a}(u,v)\big]_{m+N_x n} = \frac{1}{\sqrt{N}}\, e^{\,j\pi(m u + n v)},
\end{align}
where $m=0,\dots,N_x-1$ and $n=0,\dots,N_y-1$.
Since the occupied bandwidth is usually negligible compared to the carrier frequency in practice, the half-wavelength spacing $\lambda / 2$ is approximately the same across all carriers used by the system's DSA scheme, which Section~\ref{sec:proposed} explains in detail.

\subsection{User Partitioning}
\label{sec:partitioning}

The users partition into $B$ disjoint groups $\{\mathcal{P}_b\}_{b=1}^{B}$, where $\mathcal{P}_b$ denotes the user set served by BS $b$.
Each user is served by a single BS at any given time. 
Hence, $\mathcal{P}_b\cap\mathcal{P}_{b'}=\varnothing$ for $b\neq b'$,
and $\bigcup_{b=1}^{B}\mathcal{P}_b=\{1,\cdots,K\}$.
$b(k)$ is the unique BS serving user $k$, i.e., $k\in\mathcal{P}_{b(k)}$.
This work considers three baseline partitioning schemes:
($i$) a \emph{random} balanced partition, where $|\mathcal{P}_b| \in \{\lfloor K/B \rfloor, \lceil K/B \rceil\}$ $\forall b$,
($ii$) a \emph{greedy} balanced partition where each BS sequentially selects the nearest $|\mathcal{P}_b|$ users,
and ($iii$) the \emph{Voronoi} partition, i.e., $b(k)=\arg\!\min_{b}d_{b,k}$, where $|\mathcal{P}_b| \in \{0, 1, \cdots, K\}$ $\forall b$.
The BS is inactive if $|\mathcal{P}_b| = 0$.

\subsection{Downlink Signal Model}

The system model has each BS $b \in \{1, \cdots, B\}$ transmit a precoded superposition of the symbols $\mathbf{s}_b=[s_k]_{k\in\mathcal{P}_b} \in \mathbb{C}^{|\mathcal{P}_b| \times 1}$ that target its group $\mathcal{P}_b$, as
\begin{align}
    \label{eq:tx}
    \mathbf{x}_b = \sqrt{P_b}\sum_{k\in\mathcal{P}_b}\mathbf{w}_{b,k}\,s_k = \sqrt{P_b}\,\mathbf{W}_b\mathbf{s}_b,
\end{align}
where $\mathbf{w}_{b,k}\in\mathbb{C}^{N \times 1}$ is the precoding vector for user $k$, 
$\mathbf{W}_b=[\mathbf{w}_{b,k}]_{k\in\mathcal{P}_b} \in\mathbb{C}^{N \times |\mathcal{P}_b|}$,
$\mathrm{Tr}(\mathbf{W}_b\mathbf{W}_b^{\sf H})=1$,
$\mathbb{E}[\mathbf{s}_b\mathbf{s}_b^{\sf H}]=\mathbf{I}$,
and $P_b$ is the transmit power of BS $b$.

Operating under a DSA mechanism, each BS $b$ is assigned a carrier frequency $f_b$ and power $P_b$.
The interference between two BSs depends on their carriers' spectral overlap, captured by the coefficient $\rho(\Delta f)$.
For two BSs that share the same frequency spectrum ($\Delta f=0$), $\rho(0) = 1$.
For two BSs that use a different spectrum ($\Delta f > {\sf BW}$), where ${\sf BW}$ denotes the occupied passband bandwidth of each spectrum slot, $\rho(\Delta f) \approx 0$.
Simulations treat non-overlapping bands as orthogonal, i.e., $\rho(\Delta f) \in \{0,1\}$.
Then, the signal at receiver $k$, served by BS $b=b(k)$, is
\begin{align}
    \label{eq:rx}
    y_k =\; & \sqrt{P_b} \mathbf{h}_{b,k}^{\sf H} \mathbf{w}_{b,k} s_k
    + \sqrt{P_b} \sum_{\substack{m\in\mathcal{P}_b,\\ m\neq k}} \mathbf{h}_{b,k}^{\sf H}\mathbf{w}_{b,m} s_m \nonumber\\
    & + \sum_{j\neq b}\sqrt{\rho(f_j-f_b) P_j}\sum_{m\in\mathcal{P}_j}\mathbf{h}_{j,k}^{\sf H}\mathbf{w}_{j,m} s_m + z_k,
\end{align}
where $\mathbf{h}_{b,k}\in\mathbb{C}^{N \times 1}$ denotes the downlink channel from BS $b$ to user $k$,
$z_k\sim\mathcal{CN}(0,\sigma^2)$ denotes the additive white Gaussian noise,
and $j \in \{1, \cdots, B\}$.
The first three terms in \eqref{eq:rx} represent the desired signal, the \emph{intra-group} interference, and the \emph{inter-group} interference, respectively.

\subsection{Achievable Rate}

The SINR of user $k$, with $b=b(k)$, is
\begin{align}
    \label{eq:sinr}
    \gamma_k = \frac{P_b\,\big|\mathbf{h}_{b,k}^{\sf H}\mathbf{w}_{b,k}\big|^2}{I^{\sf intra}_k + I^{\sf inter}_k + \sigma^2},
\end{align}
where the intra- and inter-group interference powers are
\begin{align}
    \label{eq:intra}
    I^{\sf intra}_k &= P_b\!\!\sum_{\substack{m\in\mathcal{P}_b,\\ m\neq k}}\!\!\big|\mathbf{h}_{b,k}^{\sf H}\mathbf{w}_{b,m}\big|^2, \\
    \label{eq:inter}
    I^{\sf inter}_k &= \sum_{j\neq b} \rho(f_j-f_b)\,P_j\!\!\sum_{m\in\mathcal{P}_j}\!\!\big|\mathbf{h}_{j,k}^{\sf H}\mathbf{w}_{j,m}\big|^2.
\end{align}
Then, the ergodic sum spectral efficiency can be expressed as
\begin{align}
    \label{eq:sumse}
    \bar{R} = \mathbb{E}\!\left[\sum_{k=1}^{K}\log_2\!\big(1+\gamma_k\big)\right],
\end{align}
where the expectation is taken over the user locations and channel realizations.
Section~\ref{sec:proposed} addresses a joint design of the precoders $\{\mathbf{W}_b\}$ and the carrier frequency $\{f_b\}$ and power $\{P_b\}$ assignment.

\section{Proposed Cooperative DSA Scheme}
\label{sec:proposed}

The proposed cooperative DSA scheme manages user interference using two layers.
The \emph{spatial} layer designs, at each BS, a linear precoder that suppresses the intra-group interference in \eqref{eq:rx}.
The \emph{spectral deconfliction} layer assigns each BS $b$ a carrier frequency $f_b$ and transmit power $P_b$ so that the interference experienced by every user in the system is kept under a certain threshold.
The inter-group interference is considered here.

Before running the algorithm's two layers, the model incorporates a sequential admission procedure: a BS $b$ and its respective group $\mathcal{P}_b$ join the algorithm as a set one at a time.
Here, the algorithm runs the spatial layer first, where the newly admitted BS $b$ does beamforming to align the beam towards its group $\mathcal{P}_b$ and to protect the already-admitted users from unnecessary interference.
The set of protected users is the \emph{victim set}.
$\mathcal{A}$ denotes the \emph{admit set}, which is the set of BSs admitted before BS $b$.
Since BS $b$ must protect only those sharing the same carrier $f_b$, the victim set is
\begin{align}
    \label{eq:victimset}
    \mathcal{V}_b = \bigcup_{\substack{b'\in\mathcal{A},\\ \rho(f_{b'}-f_b) = 1}} \mathcal{P}_{b'}.
\end{align}

If the spatial layer is not sufficient to pass the compatibility test, the algorithm then runs the spectral deconfliction layer.
In this layer, BS $b$ shifts its carrier frequency $f_b$ to the next spectrum slot and repeats the test.
More details regarding the spatial and spectral deconfliction layers appear in Algorithm~\ref{alg:admit}.

\subsection{Iterative Water-filling via Uplink-Downlink Reciprocity}
\label{subsec:iwf}

The proposed DSA scheme incorporates the IWF technique \cite{yu2004iterative} into our cell-free multiuser downlink MISO setting and uses it as our main precoding method.
The IWF technique is well-suited for DSA because it provides a good balance of interference mitigation, power optimization, and spectral efficiency performance.
Since the IWF technique~\cite{yu2004iterative} considers the uplink setting, uplink-downlink reciprocity~\cite{rashid1998transmit} transforms the downlink problem into a virtual uplink problem.
Perfect channel state information at the transmitter (CSIT) is assumed throughout this paper.

Consider BS $b$ serving its group $\mathcal{P}_b$.
Deferring the mitigation of the inter-group interference in \eqref{eq:rx} to the spectral deconfliction layer, 
the channel serving the intra-group is a MISO downlink channel under the per-BS transmit power budget $P_b$.
Reciprocity gives a virtual uplink using the same channel vectors: a single-input multiple-output (SIMO) multiple access channel (MAC) where each user $k\in\mathcal{P}_b$ transmits over the conjugated channel $\mathbf{h}_{b,k}$ with power $q_{b,k}$ under the constraint $\sum_{k\in\mathcal{P}_b} q_{b,k}\le P_b$.
Reusing the symbols $s_m$ of \eqref{eq:tx}, now transmitted by the users rather than by BS $b$, the signal received at BS $b$ in the virtual uplink is
\begin{align}
    \label{eq:macrx}
    \mathbf{y}_b^{\sf MAC} 
    &= \sum_{m\in\mathcal{P}_b}\sqrt{q_{b,m}}\,\mathbf{h}_{b,m}\,s_m + \mathbf{z},
\end{align}
where $\mathbf{z} \sim \mathcal{CN}(\mathbf{0}, \sigma^2\mathbf{I}_N)$.
The interference-plus-noise covariance observed at BS $b$ when detecting user $k$'s signal is
\begin{align}
    \label{eq:Rbk}
    \mathbf{R}_{b,k} = \sigma^2\mathbf{I}_N 
    + \sum_{\substack{m\in\mathcal{P}_b,\\ m\neq k}} q_{b,m} \mathbf{h}_{b,m}\mathbf{h}_{b,m}^{\sf H} 
    + \zeta P_{\sf max} \sum_{j\in\mathcal{V}_b} \mathbf{h}_{b,j}\mathbf{h}_{b,j}^{\sf H}.
\end{align}
The last term is the leakage-aware regularization term, where $\zeta\ge0$ is a weight controlling how hard BS $b$ steers the beam away from the users in its victim set.
Adding user $k$'s own term to \eqref{eq:Rbk} gives the full covariance
$\mathbf{R}_{b,k}+q_{b,k}\mathbf{h}_{b,k}\mathbf{h}_{b,k}^{\sf H}
= \mathbf{R}_{b,0}+\sum_{m\in\mathcal{P}_b}q_{b,m}\mathbf{h}_{b,m}\mathbf{h}_{b,m}^{\sf H}$, 
where $\mathbf{R}_{b,0}$ is the effective noise term in \eqref{eq:Rbk}.
The virtual-uplink sum rate is the mutual information of this Gaussian vector channel, i.e., the log-determinant of the received covariance whitened by the effective noise \cite{jindal2005sum}:
\begin{align}
    \label{eq:macsum}
    R_b(\mathbf{q}_b) 
    = \log_2 \! \bigg| \mathbf{R}_{b,0} \! + \!\!\!
    \sum_{m\in\mathcal{P}_b} \!\! q_{b,m}\mathbf{h}_{b,m}\mathbf{h}_{b,m}^{\sf H} \bigg|
    \! - \! \log_2 \! \big|\mathbf{R}_{b,0}\big|,
\end{align}
where $\mathbf{q}_b = [q_{b,k}]_{k\in\mathcal{P}_b}$.
For simplicity, a local index $k' \in \{ 1, \cdots, |\mathcal{P}_b| \}$ enumerates the users $k\in\mathcal{P}_b$.
Then, iteratively using the matrix determinant lemma $|\mathbf{A}+\mathbf{v}\mathbf{v}^{\sf H}|=|\mathbf{A}|(1+\mathbf{v}^{\sf H}\mathbf{A}^{-1}\mathbf{v})$ 
for invertible $\mathbf{A}$, the sum rate in \eqref{eq:macsum} becomes
\begin{align}
    \label{eq:decomp}
    R_b(\mathbf{q}_b)
    = \sum_{k'=1}^{|\mathcal{P}_b|} \log_2 \big(1+q_{b,(k')}\nu_{b,(k')}\big),
\end{align}
where
\begin{align}
    \label{eq:nu}
    \nu_{b,(k')} 
    = \mathbf{h}_{b,(k')}^{\sf H} \!\!
        \left( \mathbf{R}_{b,0} \! + \! \sum_{j=1}^{k'-1} q_{b, (j)} \mathbf{h}_{b,(j)}\mathbf{h}_{b,(j)}^{\sf H} \! \right)^{\!\!\! -1} \!\!\!
            \mathbf{h}_{b,(k')}.
\end{align}
The matrix inside the inversion operator in \eqref{eq:nu} is invertible for all $(k')$ because it is a Hermitian positive definite matrix.

Maximization of the virtual-uplink sum rate in \eqref{eq:decomp} for $q_{b,(k')}\ge0$ and $\sum_{k'=1}^{|\mathcal{P}_b|}q_{b,(k')}\le P_b$ uses the IWF technique \cite{yu2004iterative}.
Fixing $\nu_{b,(k')}$ for now, the water-filling solution is
\begin{align}
    \label{eq:wf}
    q_{b,(k')} = \Big[\mu_b - \tfrac{1}{\nu_{b,(k')}}\Big]^{+}, \qquad 
    \sum_{k' = 1}^{|\mathcal{P}_b|} q_{b,(k')} = P_b,
\end{align}
where $[\cdot]^{+}=\max(\cdot,0)$ and $\mu_b$ denotes the water level.
The solution in \eqref{eq:wf} satisfies the Karush-Kuhn-Tucker conditions of this maximization with $\nu_{b,(k')}$ held fixed.
Meanwhile, since $\nu_{b,(k')}$ in \eqref{eq:nu} depends on the powers $q_{b,(k')}$ in \eqref{eq:wf} and vice versa,
the algorithm iteratively computes $\nu_{b,(k')}$ and $q_{b,(k')}$ in an alternating fashion.
At iteration $n$, it uses $\mathbf{q}_b^{(n-1)}$ from the previous iteration to compute $\{\nu_{b,(k')}\}_{k'=1}^{|\mathcal{P}_b|}$ in \eqref{eq:nu}, then it computes $\tilde{\mathbf{q}}_b^{(n)}$ via water-filling in \eqref{eq:wf}, and lastly the update is weighted-averaged as follows \cite{jindal2005sum}:
\begin{align}
    \label{eq:damped}
    q_{b,(k')}^{(n)} = \tfrac{1}{|\mathcal{P}_b|}\,\tilde{q}_{b,(k')}^{(n)} + \tfrac{|\mathcal{P}_b|-1}{|\mathcal{P}_b|}\,q_{b,(k')}^{(n-1)}.
\end{align}
Initialization is $q_{b,(k')}^{(0)}=P_b/|\mathcal{P}_b|$ for all $k'$ and the algorithm stops the iteration once $\|\mathbf{q}_b^{(n)}-\mathbf{q}_b^{(n-1)}\|_2 \leq \epsilon P_b$ is true.
The overall procedure of the IWF algorithm, including the iterative method of determining $\mu_b$, is outlined in Algorithm~\ref{alg:waterfill}.
Algorithm~\ref{alg:waterfill} converged within $n_{\max}$ iterations in all simulations.

\begin{algorithm} [t]
\caption{Iterative Water-Filling (IWF)} \label{alg:waterfill}
Initialize $q_{b,(k')}^{(0)} \leftarrow P_b/|\mathcal{P}_b|$ for all $k'$\\
\For{$n = 1, 2, \cdots, n_{\max}$}{
Compute $\{\nu_{b,(k')}\}_{k'=1}^{|\mathcal{P}_b|}$ in \eqref{eq:nu} using $\mathbf{q}_b^{(n-1)}$\\
Define $a_{(k')} \gets 1/\nu_{b,(k')}$\\
Sort ascending as $a_{<1>}\le\cdots\le a_{<|\mathcal{P}_b|>}$\\
\For{$m = |\mathcal{P}_b|, |\mathcal{P}_b|-1, \cdots, 2, 1$}{
$\mu_b \gets \big(P_b+\textstyle\sum_{\ell\le m} a_{<\ell>}\big)/m$\\
\If{$\mu_b > a_{<m>}$}{
\textbf{break}
}
}
Water-fill $\tilde{q}_{b,(k')}^{(n)} \gets \big[\mu_b - a_{(k')}\big]^{+}$ as in \eqref{eq:wf}\\
Average the update as in \eqref{eq:damped}\\
\If{$\|\mathbf{q}_b^{(n)}-\mathbf{q}_b^{(n-1)}\|_2 \le \epsilon P_b$}{
\textbf{break}
}
}
\Return{\ }{$\mathbf{q}_b^{(n)}$}
\end{algorithm}

\subsection{Precoding via Uplink-Downlink Reciprocity}
\label{sec:precoding}

At convergence, user $k$'s virtual-uplink receive filter is the linear minimum mean-square-error filter \cite{van2002optimum, bjornson2014optimal}, and the same vector serves as its downlink beam direction:
\begin{align}
    \label{eq:mmse}
    \hat{\mathbf{u}}_{b,(k')} = \frac{ \big( \mathbf{R}_{b,(k')}^\star \big)^{-1}\mathbf{h}_{b,(k')}}{\big\|\big( \mathbf{R}_{b,(k')}^\star \big)^{-1}\mathbf{h}_{b,(k')}\big\|},
\end{align}
where $\mathbf{R}_{b,(k')}^\star$ is the interference-plus-noise covariance in \eqref{eq:Rbk} using the converged powers $\mathbf{q}_b^\star$.
The algorithm sets the relative power allocated to user $(k')$ as $p_{b,(k')}$, which the power control in Section~\ref{subsec:power} determines:
\begin{align}
    \label{eq:assemble}
    \mathbf{F}_b
    = \left[
        \sqrt{p_{b,(1)}}\,\hat{\mathbf{u}}_{b,(1)}, \cdots, \sqrt{p_{b,(|\mathcal{P}_b|)}}\,\hat{\mathbf{u}}_{b,(|\mathcal{P}_b|)}
    \right].
\end{align}
Then, the precoder of BS $b$ is normalized to unit total power:
\begin{align}
    \label{eq:W}
    \mathbf{W}_b = \frac{\mathbf{F}_b}{\sqrt{\mathrm{Tr}(\mathbf{F}_b\mathbf{F}_b^{\sf H})}}.
\end{align}
This is the \emph{IWF precoder}.
Using the covariance in \eqref{eq:Rbk}, the IWF precoder nulls harder toward other users that receive large interference.
This paper also uses two simple precoders as baselines: maximum ratio transmission (MRT) and regularized zero-forcing (RZF).
MRT ($\mathbf{W}_b\propto\mathbf{H}_b$, $\mathbf{H}_b=[\mathbf{h}_{b,k}]_{k\in\mathcal{P}_b}$) does not consider the interference and prioritizes the users with the best channels,
whereas RZF ($\mathbf{W}_b\propto\mathbf{H}_{b}\big(\mathbf{H}_{b}^{\sf H}\mathbf{H}_{b}+\tfrac{|\mathcal{P}_b|\sigma^2}{P_b}\mathbf{I}\big)^{-1}$) prioritizes suppressing the intra-group interference.
Both baselines are normalized to unit power as in \eqref{eq:W}, and each BS scales its transmit power $P_b$, up to the maximum transmit power $P_{\sf max}$, such that its user with the weakest beamforming gain $|\mathbf{h}_{b,k}^{\sf H}\mathbf{w}_{b,k}|^2$ achieves the target SINR $\gamma_{\sf tgt}$.

\subsection{Downlink Power Allocation and Power Control}
\label{subsec:power}

The proposed DSA mechanism uses a power-control method that limits each BS to use only as much transmit power $P_b$ as its users require to reach the target SINR $\gamma_{\sf tgt}$, up to $P_{\sf max}$:
\begin{align}
    & p_{b,(k')} = \frac{\gamma_{\sf tgt} (\sigma^2 + I_{\sf Rx})}{|\mathbf{h}_{b,(k')}^{\sf H}\hat{\mathbf{u}}_{b,(k')}|^2}, \\
    \label{eq:power}
    & P_b = \min \left( P_{\sf max},
    \sum_{k'=1}^{|\mathcal{P}_b|} p_{b,(k')} \right),
\end{align}
where $I_{\sf Rx} = \sigma^2\,\mathrm{INR}_{\sf tgt}$ is the largest aggregate interference that the SCM underlay mask of Section~\ref{subsec:scm} allows,
and $\mathrm{INR}_{\sf tgt}$ is the target interference-to-noise ratio.
For instance, $\mathrm{INR}_{\sf tgt}=0$~dB allows an aggregate interference equal to the noise floor.
The value of $\gamma_{\sf tgt}$ depends on the system requirements.
By limiting each BS's transmit power, the interference that leaks into neighboring groups is reduced.

\subsection{SCM Compatibility Test}
\label{subsec:scm}

The compatibility test (CT) in the IEEE~1900.5.2 standard determines whether the spectrum can be shared by two or more RF devices.
A spectrum mask $M_{\sf Tx}(\Delta f)$, which is used to describe a BS's relative emitted power at offset $\Delta f$ from its carrier frequency, appears in Fig.~\ref{fig:mask_tx}.
An underlay mask $M_{\sf Rx}(\Delta f)$ describes the relative interference a user can tolerate, as in Fig.~\ref{fig:mask_rx}.
The masks in Fig.~\ref{fig:masks} are derived from measurements of binary phase shift keying (BPSK) modulated signals with a $1$~MHz bandwidth in \protect\cite{stojadinovic2022spectrum}.

The interference power that BS $j$ delivers to a victim user $k_{\sf v}$ in BS $b_{\sf v}$'s group is $I_{j\to k_{\sf v}} = P_j \big\|\mathbf{W}_j^{\sf H} \mathbf{h}_{j,k_{\sf v}}\big\|_2^2$.
If BS $b$ is the newly admitted BS, a victim user $k_{\sf v} \in \mathcal{P}_{ \mathcal{A}\cup\{b\} }$ is compatible if, at every frequency $f$, the received interference power stays below the allowed threshold:
\begin{align}
    \label{eq:ct}
    I^{\sf intra}_{k_{\sf v}} \, M_{\sf Tx}(f-f_{b_{\sf v}})
    & + \!\!\!\sum_{\substack{j\in\mathcal{A}\cup\{b\},\\ j\neq b_{\sf v}}}\!\!\! I_{j\to k_{\sf v}} \, M_{\sf Tx}(f-f_j) \\
    \nonumber
    & \quad \quad \quad \quad \quad \quad  \leq I_{\sf Rx} \, M_{\sf Rx}(f-f_{b_{\sf v}}),
\end{align}
where $I^{\sf intra}_{k_{\sf v}}$ is computed using \eqref{eq:intra} and $I_{\sf Rx}$ is defined in Section~\ref{subsec:power}.
The CT passes only if all admitted users in the sequential admission process are compatible.
If the CT fails even on a fresh spectrum slot, the failure is due to intra-group interference.
In this case, the BS is still admitted on that slot.
Algorithm~\ref{alg:admit} summarizes the overall procedure of the DSA algorithm.

\begin{algorithm} [t]
\caption{Sequential DSA Admission} \label{alg:admit}
Initialize admit set $\mathcal{A}\gets\varnothing$\\
\For{$b = 1, 2, \cdots, B$}{
\For{$c = 1, 2, \cdots, B$}{
Initialize transmit power $P_b = P_{\sf max}$\\
Select spectrum slot $f_b \gets f_{\sf ref} + (c-1)\Delta$\\
Define victim set $\mathcal{V}_b$ as in \eqref{eq:victimset}\\
Build $\{\hat{\mathbf{u}}_{b,(k')}\}$ in \eqref{eq:mmse} using Algorithm~\ref{alg:waterfill}\\
Set $P_b$ and $\mathbf{W}_b$ as in \eqref{eq:power} and \eqref{eq:W}\\
\If{$\mathrm{CT}(\mathcal{A}\cup\{b\})$ in \eqref{eq:ct} passes or no BS in $\mathcal{A}$ uses $f_b$}{
Update admit set $\mathcal{A} \gets \mathcal{A}\cup\{b\}$\\
\textbf{break}
}
}
}
\Return{\ }{$\{f_b\}$, $\{P_b\}$, $\{\mathbf{W}_b\}$} 
\end{algorithm}

\section{Numerical Results}

The downlink channel $\mathbf{h}_{b,k}$ follows a Rician model comprising a deterministic line-of-sight (LoS) component along $\mathbf{a}(u_{b,k},v_{b,k})$ in \eqref{eq:steering} and a non-line-of-sight (NLoS) component:
\begin{align}
    \label{eq:channel}
    \mathbf{h}_{b,k} = \sqrt{\beta_{b,k}} \left(
        \sqrt{\tfrac{\kappa N}{\kappa+1}} \mathbf{a}(u_{b,k},v_{b,k}) + \sqrt{\tfrac{1}{\kappa+1}}\,\mathbf{g}_{b,k}
    \right),
\end{align}
where $\kappa$ is the Rician $K$-factor, $\mathbf{g}_{b,k}\sim\mathcal{CN}(\mathbf{0},\mathbf{I}_N)$ is the NLoS term, and $\beta_{b,k}$ is the per-antenna large-scale gain, so that $\mathbb{E}\!\left[\|\mathbf{h}_{b,k}\|^2\right]=N\beta_{b,k}$.
The large-scale gain follows the $28$~GHz close-in path loss model \cite{sun2016propagation}:
\begin{align}
    \label{eq:pathloss}
    \beta_{b,k}\,[\text{dB}] = -\!\left( \mathrm{PL}_0 + 10\,\eta\log_{10}\!\frac{d_{b,k}}{d_0} + \chi_{b,k}\right),
\end{align}
with free space path loss $\mathrm{PL}_0$ at distance $d_0$, path loss exponent $\eta$, and log-normal shadowing $\chi_{b,k}\sim\mathcal{N}(0,\sigma_{\sf sf}^2)$.

Unless specified otherwise, the system parameters are as follows:
$B=16$ ceiling-mounted BSs at height $h_b=10$~m,
$K=64$ users,
$L=50$~m coverage-area side length,
$h_{\sf u}=1.5$~m user height,
$N=N_xN_y=64$ array elements with $N_x=N_y=8$,
$f_{\sf ref} = 28$~GHz default carrier frequency,
${\sf BW}=1$~MHz,
and $7$~dB receiver noise figure.
The path loss exponent $\eta=2.15$ and the shadowing standard deviation $\sigma_{\sf sf}=4.3$~dB are those of the indoor-factory (InF) LoS model in \cite{3gpp38901}.
$\mathrm{PL}_0$ at reference distance $d_0=1$~m is $\mathrm{PL}_0=20\log_{10}(4\pi d_0/\lambda)=61.4$~dB \cite{sun2016propagation}.
The Rician $K$-factor is set to $\kappa=10$.
The DSA algorithm parameters are as follows:
$P_{\sf max}=20$~dBm,
$\gamma_{\sf tgt} = 6$~dB,
$\mathrm{INR}_{\sf tgt}=6$~dB,
$\zeta=10^{-5}$,
and $\Delta = 2$~MHz.
The algorithm parameters of Algorithm~\ref{alg:waterfill} are $\epsilon=10^{-6}$ and $n_{\max}=50$.
The BSs are admitted in the order $b=1,\cdots,B$, though the BS admit order did not impact the performance in the simulations.
The default partitioning scheme is the Voronoi partition and the default precoding method is the IWF precoder.
Results are averaged over $100$ independent Monte Carlo trials.

\begin{figure}[!t]
    \centering
    \includegraphics[width=0.9\columnwidth]{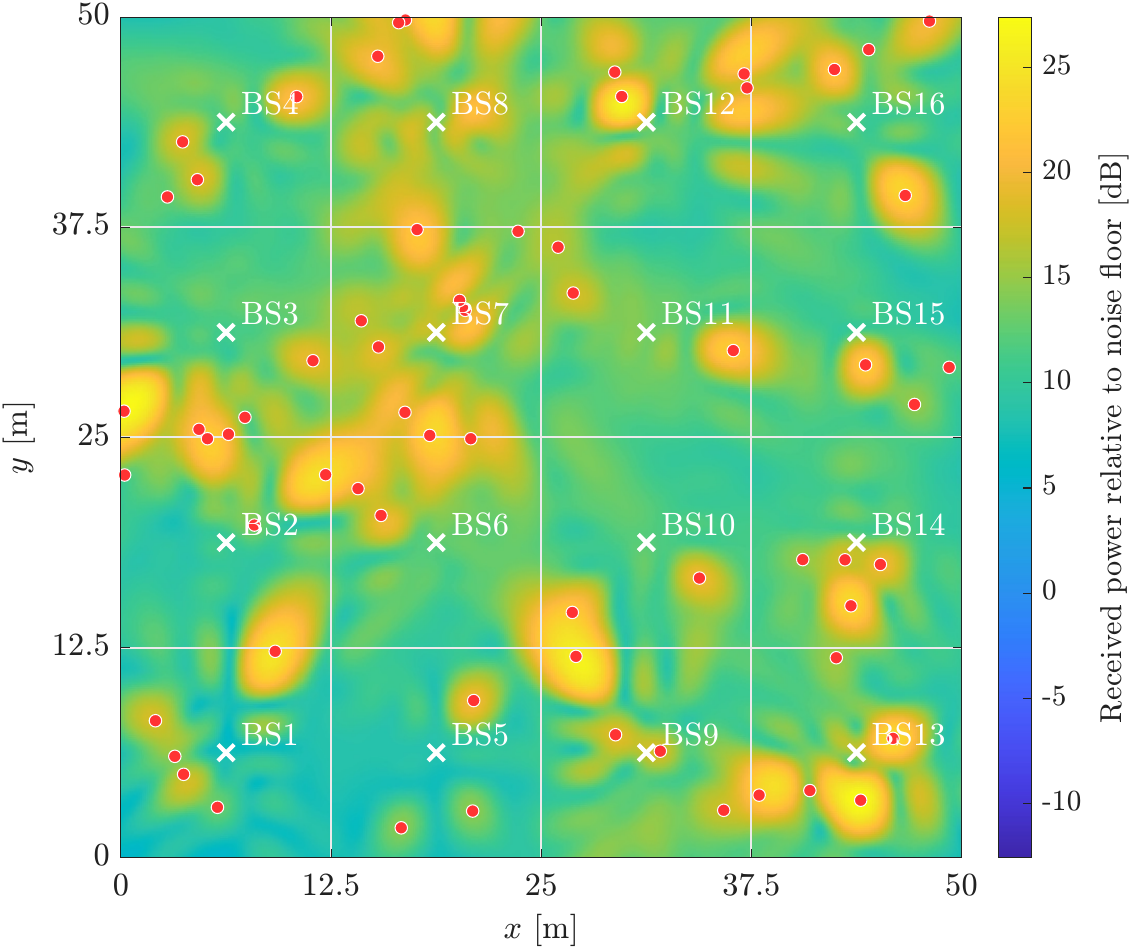}
    \caption{
        Received power over the horizontal plane at the user height $h_{\sf u}$, relative to the noise floor $\sigma^2$, after a single run of Algorithm~\ref{alg:admit} using the Voronoi partition.
        The red markers denote the $K=64$ users, the white crosses denote the $B=16$ ceiling-mounted base stations, and the white lines indicate the Voronoi regions.
    }
    \label{fig:power_map}
\end{figure}

Fig.~\ref{fig:power_map} maps the power relative to the noise floor $\sigma^2$ that the $B$ BSs deliver to the horizontal plane at the user height $h_{\sf u}$ across the service area.
The value at each position of that plane is the power delivered by every beam of every BS after a single run of the IWF precoder using the Voronoi partition, regardless of the carrier frequency.
As shown in Fig.~\ref{fig:power_map}, each BS utilizes the spatial dimension to concentrate its emission on the users of its own group, which facilitates spectrum sharing.

\begin{figure}[!t]
    \centering
    \subfigure[Spectrum use]{%
          \includegraphics[width=0.49\columnwidth]{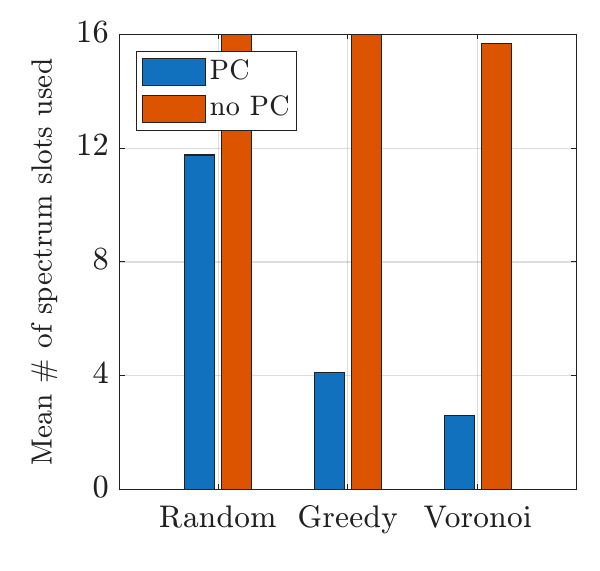}%
          \label{fig:pc_1}}
    \hfil
    \subfigure[Spectrum slot efficiency]{%
          \includegraphics[width=0.49\columnwidth]{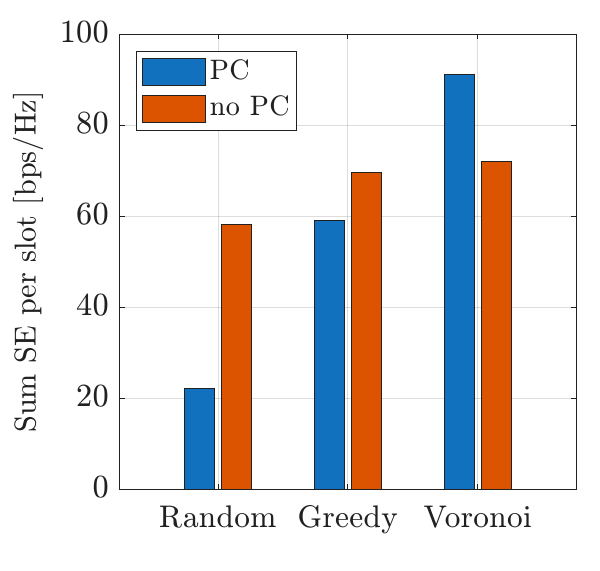}%
          \label{fig:pc_2}}
    \caption{
        The impact of the power control (PC) in \eqref{eq:power} on (a) the mean number of spectrum slots that the $B=16$ base stations occupy and (b) the ergodic sum spectral efficiency (SE) per occupied slot, for the three partitioning schemes introduced in Section~\ref{sec:partitioning}.
    }
    \label{fig:pc}
\end{figure}

Fig.~\ref{fig:pc} shows the effect of the proposed power control in \eqref{eq:power} using the three partitioning schemes introduced in Section~\ref{sec:partitioning}: random, greedy, and Voronoi partitions.
Fig.~\ref{fig:pc_1} plots the mean number of spectrum slots used with or without power control (PC).
Without PC, the $B=16$ BSs use $P_{\sf max}$ and occupy nearly all the available spectrum slots with no reuse, whereas PC reduces the count to roughly $2.6$ slots using the Voronoi partition.
This is because PC allows better spectrum sharing by reducing unnecessary interference inflicted toward other users while providing sufficient SINR.
However, PC comes at the cost of the spectral efficiency performance, since it prioritizes spectrum sharing over using $P_{\sf max}$.
Fig.~\ref{fig:pc_2} plots the ergodic sum spectral efficiency (SE) per occupied spectrum slot, with or without PC.
The Voronoi partition is the only partitioning scheme that has a better sum SE per slot performance with PC, whereas the random and greedy partitions lose more throughput than the spectrum slots they save.
Nevertheless, PC is useful in scenarios where the spectrum is limited and the RF devices do not require high spectral efficiency performance.

\begin{figure}[!t]
    \centering
    \subfigure[Spectrum use]{%
          \includegraphics[width=0.49\columnwidth]{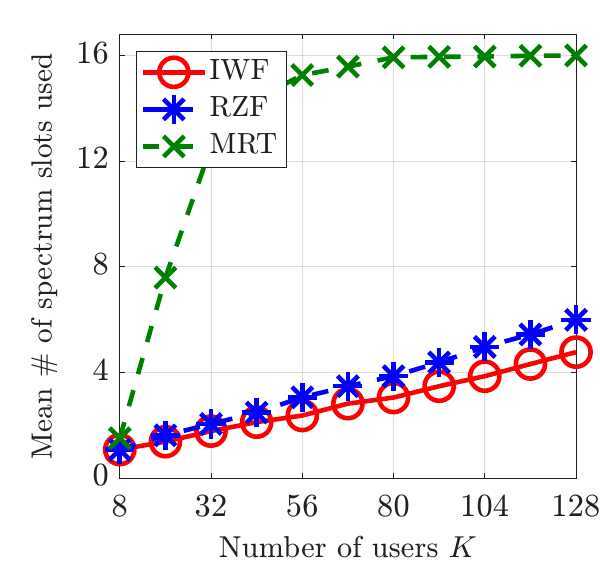}%
          \label{fig:sweep_K_1}}
    \hfil
    \subfigure[Spectrum slot efficiency]{%
          \includegraphics[width=0.49\columnwidth]{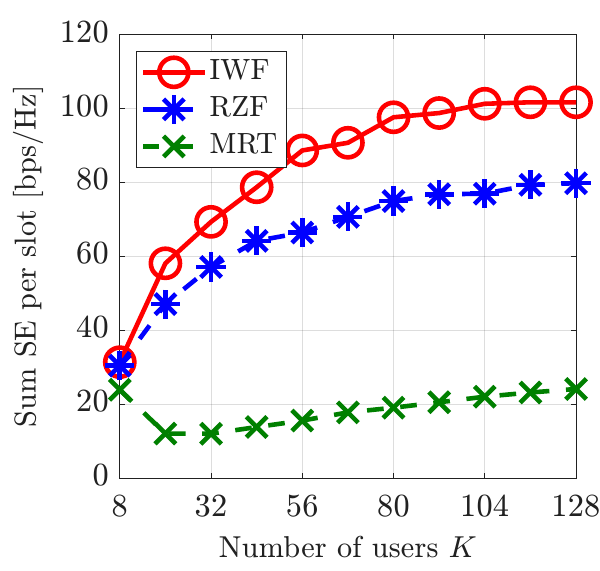}%
          \label{fig:sweep_K_2}}
    \caption{
        Comparison of the precoding methods IWF, RZF, and MRT with respect to the number of users $K \in \{ 8, 20, \cdots, 128 \}$.
        (a) shows the mean number of spectrum slots that the $B=16$ base stations occupy and (b) shows the ergodic sum spectral efficiency (SE) per occupied slot.
    }
    \label{fig:sweep_K}
\end{figure}

Fig.~\ref{fig:sweep_K} compares the IWF precoder with the two simple baseline precoders MRT and RZF introduced in Section~\ref{sec:precoding} for $K \in \{ 8, 20, \cdots, 128 \}$.
Fig.~\ref{fig:sweep_K_1} plots the mean number of occupied spectrum slots.
The MRT precoder forces the BSs to use more spectrum slots because it does not consider the intra-group interference or the inter-group interference.
The RZF and IWF precoders, on the other hand, use far fewer spectrum slots because they steer the beams to reduce the intra-group interference.
In the case of the IWF precoder, it also considers the inter-group interference via the leakage-aware regularization term in \eqref{eq:Rbk}.
This plot shows that \emph{how} the spatial dimension is utilized can have a huge impact on spectrum sharing, and hence shows the importance of placing the precoding design inside the DSA mechanism.
Fig.~\ref{fig:sweep_K_2} shows that the IWF precoder achieves the highest sum SE per occupied slot at every $K$.

\section{Conclusion}

This paper introduced a cooperative SCM-based DSA mechanism that employs uniform planar arrays at the transmitter.
The proposed DSA jointly designs the precoder and spectrum allocation to better utilize the spatial dimension for spectrum sharing.
The key takeaway is that better utilizing the spatial dimension by jointly designing the precoder and spectrum allocation is critical for spectrum sharing.
Future research directions include better partitioning schemes, nonlinear precoding, and considering the Doppler shift of moving users for localization, channel estimation, and predictive beamforming.

\section*{Acknowledgment}

This research was funded through support and collaboration from Stanford, Ericsson, and Samsung.

\bibliographystyle{IEEEtran}
\bibliography{references}

\end{document}